\documentclass{revtex4-2}

\usepackage{amssymb}
\usepackage{amsfonts}
\usepackage{amsmath}
\usepackage{graphicx}
\usepackage{subfig}

\begin{document}


\title{Position-dependent mass Schr\"{o}dinger particles in space-like screw dislocation: associated degeneracies and magnetic and Aharonov-Bohm flux fields effects}
\author{Zeinab Algadhi}
\email{zalgadhi@gmail.com}
\affiliation{Department of Physics, Bani Walid University, Bani Walid, Libya.}
\author{Omar Mustafa}
\email{omar.mustafa@emu.edu.tr}
\affiliation{Department of Physics, Eastern Mediterranean University, G. Magusa, north
Cyprus, Mersin 10 - Turkey.}

\begin{abstract}
\textbf{Abstract:} We consider non-relativistic position-dependent mass (PDM) Sch\"{o}dinger particles moving in an elastic medium with space-like screw dislocation. Within the cylindrical coordinates $\left( r,\phi ,z\right) $, we study and report the effects of screw dislocation as well as PDM settings on the energy levels of some PDM-quantum mechanical systems. In so doing, we use a power-law type positive-valued dimensionless scalar multiplier $%
f(r)=\lambda r^{\sigma }$ (which manifestly introduces the metaphoric notion of PDM-Sch\"{o}dinger particles). Next, we subject such PDM particles to magnetic and Aharonov-Bohm flux fields. We report exact or conditionally exact eigenvalues and eigenfunctions for $V(r)=0$ and $V(r) =a+b\,r+c\,r^{2}$.


\end{abstract}

\maketitle

\section{Introduction}

The influence of topological defects on the physical properties of various physics systems has been widely discussed in the literature \cite{1,2,3,4}. In gravitation \cite{5,6,7}, in condensed matter physics \cite{8,9,10}, cosmic string \cite{11,12}, and disclinations or dislocations in solids and liquid crystals 
\cite{9,13,14}. Based on the geometric theory \cite{15,16}, topological defects cause elastic deformations in continuous media, which are described by a line element and metric. In particular, a Riemann-Cartan manifold \cite{17}, defined dislocations and disclinations as topological defects in the elastic medium. Dislocations and disclinations are associated with torsion and curvature, respectively.

On the other hand, the concept of position-dependent mass (PDM in short) has sparked research interest in both classical and quantum mechanics 
\cite{18,19,20,21,22,23,24,25,26,27,28,29,30,31,32,33,34,35,36,37,38,39,40}. PDM is a metaphoric manifestation of coordinate transformation \cite{22,23,24,28}. PDM settings manifestly change the form of the canonical momentum in classical (i.e., $\textbf{p}(\textbf{r})=M(r)\,\dot{\textbf{r}}=m_\circ\,f(\textbf{r})\,\dot{\textbf{r}}$, where $f(\textbf{r})$ is a positive valued dimensionless scalar multiplier) and the momentum operator in quantum mechanics (see (\ref{eq:P}), below).

Several authors have investigated the effects of screw dislocation on quantum systems \cite{5,37,38,39,40,41,42,43,44}. Aharonov- Bohm effects with or without screw dislocations have been discussed in different systems by some others \cite{45,46,47,48,49}. Only a handful number of attempts were made to treat PDM quantum systems under these effects (c.f., e.g.,  \cite{50,51,52} and references cited therein). Very recently, Mustafa \cite{50,51,52} has investigated the relativistic PDM quantum system (using Klein- Gordon equation) in the presence of the magnetic and Aharonov-Bohm flux fields and under the influence of the different topological defects. To the best of our knowledge, however, studies of the topological defects on the non-relativistic quantum system with PDM have  not yet been explored/investigated. To fill this gap, we discuss non-relativistic quantum systems for PDM Schr\"{o}dinger particles in the presence of space-like screw dislocation and magnetic and Aharonov-Bohm flux fields.

In the current methodical proposal, we shall follow the von-Roos \cite{19} PDM kinetic energy operator (in $\hbar =2m_{\circ}=1$ units) and use Mustafa and Mazharimousavi's ordering \cite{25} ( known in the literature as MM-ordering) to cast the kinetic energy operator as %
\begin{equation}  \label{eq:T}
\widehat{T}=\left( \frac{\widehat{p}\left( \textbf{r}\right) }{\sqrt{%
f\left( \textbf{r}\right) }}\right) ^{2}= - f\left( \textbf{r}%
\right) ^{-1/4} \nabla f\left(\textbf{ r}\right) ^{-1/2}\ \nabla
f\left(\textbf{r}\right) ^{-1/4},
\end{equation}%
where $f\left( \textbf{r}\right)$ is a positive valued dimensionless scalar multiplier that forms the position-dependent mass $M\left( \textbf{r}\right) =m_{\circ }f\left( \textbf{r}\right)$ (in the von-Roos Hamiltonian, with $m_{\circ }$ as the rest mass), and $\widehat{p}\left(\textbf{r}\right)$ is the PDM-momentum operator constructed by Mustafa and Algadhi \cite{27}, to read%
\begin{equation}  
\widehat{P}\left( \textbf{r}\right) =-i\left[\mathbf{\nabla}-\frac{1}{4}\left( \frac{\mathbf{\nabla}f\left( \textbf{r}\right) }{f\left( \textbf{r}\right) }\right) \right]. \label{eq:P}
\end{equation}%
The reader is advised to refer to Mustafa and Mazharimousavi \cite{25} for comprehensive  details on the strict determination of the powers of $f\left( \textbf{r}\right) $ and refer to Mustafa and Algadhi \cite{27} for the structure of the PDM-momentum operator. We use this result (\ref{eq:P}) to describe PDM quantum particles throughout.

The present paper is organized as follows. In section II, we consider an elastic medium with a screw dislocation, and then, we introduce the topological defect of the screw dislocation for the PDM particle within the cylindrical coordinates $\left( r,\phi ,z\right) $. Then, by considering the PDM kinetic operator (\ref{eq:T}), and the PDM momentum operator  (\ref{eq:P}). We derive the PDM Hamiltonian that describes this system and their non-relativistic PDM in the background of the screw dislocation. In the same section, we obtain exact eigenvalues and eigenfunctions by using the radial power-law PDM settings for two configurations: (i)  Quasi-free PDM Schr\"{o}dinger particles with space-like screw dislocation, and (ii) PDM Schr\"{o}dinger particles under the influence of space-like screw dislocation and interaction potential. We investigate, in section III, the influence of the screw dislocation on the PDM-charged particle in the presence of magnetic and Aharonov-Bohm flux fields, and report exact or conditionally exact eigenvalues and corresponding wave functions for the two cases: without and with the confining potential by using the same PDM settings. Finally, our conclusion is given in section IV.

\section{PDM-Schr\"{o}dinger particles under the effect of space-like screw dislocation}

In this section, we start our study by discussing the effect of a space-like screw dislocation on the PDM quantum particle as one of the linear topological defects in an elastic medium. Where the screw dislocation is described in cylindrical coordinates by the following line element \cite{43,44}
\begin{equation}  \label{eq:SC}
ds^{2}=dr^{2}+r^{2}d\phi ^{2}+\left( dz+\chi \, d\phi \right) ^{2}
\end{equation}%
where $\chi$ is a parameter related to Burger's vector $b$ through $\chi =%
\frac{b}{2\pi}$.This screw dislocation (\ref{eq:SC}) presents a torsion field but no curvature. The torsion associated with this topological defect corresponds to a conical singularity at the origin \cite{53}.

The covariant and contravariant metric tensors associated with the line element (\ref{eq:SC}), respectively, read as%
\begin{equation}  \label{eq:g}
g_{ij}=\left( 
\begin{array}{ccc}
1 & 0 & 0 \\ 
0 & (r^{2}+\chi ^{2}) & \chi \\ 
0 & \chi & 1%
\end{array}%
\right) \iff  g^{ij}=\left( 
\begin{array}{ccc}
1 & 0 & 0 \\ 
0 & \frac{1}{r^{2}} & -\frac{\chi }{r^{2}} \\ 
0 & -\frac{\chi }{r^{2}} & (1+\frac{\chi ^{2}}{r^{2}})
\end{array}%
\right) ,
\end{equation}%
where $det(g)=r^2$. By considering the PDM kinetic operator in (\ref{eq:T}) and the PDM-momentum operator in \eqref{eq:P}, the Laplace-Beltrami operator with topological defects that corresponds to the metric $g^{ij}$ (as mentioned in \cite{45}) can be modified to describe the PDM Hamiltonian (in $\hbar
=2m_{\circ }=1$ units) as%
\begin{equation}  \label{eq:H1}
\hat{H}=\frac{1}{\sqrt{g}}\left( \frac{\hat{p}_{i}}{\sqrt{f\left( \mathbf{r}%
\right) }}\right) \sqrt{g}\,g^{ij}\left( \frac{\hat{p}_{j}}{\sqrt{f\left( \mathbf{r}%
\right) }}\right)
\end{equation}%
In what follows, however, we shall use the assumption that the positive valued scalar multiplier $f(\mathbf{r})=f(r)$ (i.e., only radially dependent). In this way, the non-relativistic PDM-Schr\"{o}dinger equation in the presence of the screw dislocation (\ref{eq:SC} )is given by%
\begin{eqnarray} 
&-&\frac{1}{f( r) }\left[ \partial _{r}^{2}+\frac{1}{r}\partial_{r} -\frac{f^{\prime }( r)}{f(r)}\partial _{r} 
\right. \nonumber \\
&& \left.
+\frac{1}{r^{2}}\left( \partial _{\phi }-\chi \, \partial _{z}\right)^{2}+\partial _{z}^{2}+M\left( r\right) \right] \psi +V\left( r\right) 
\psi =E\psi,  \label{eq:m}
\end{eqnarray}%
where 
\begin{equation}  \label{eq:M}
M\left( r\right) =\frac{7}{16}\left( \frac{f^{\prime }\left( r\right) }{%
f\left( r\right) }\right) ^{2}-\frac{1}{4}\left( \frac{f^{\prime \prime
}\left( r\right) }{f\left( r\right) }+\frac{f^{\prime }\left( r\right) }{%
r\, f\left( r\right) }\right),
\end{equation}%
the solution of (\ref{eq:m}) can be obtained by

\begin{equation}  \label{eq:Y}
\psi \left( r,\phi ,z\right) =e^{i\ell \phi }e^{ikz}R\left( r\right)
\end{equation}

where $k$ is a constant and the magnetic quantum number $\ell =0,\pm 1,\pm 2,\cdots$. Thus, the radial equation of (\ref{eq:m}) is obtained as

\begin{eqnarray}  \label{eq:R1}
&&\frac{R^{\prime \prime }\left( r\right) }{R\left( r\right) }+\left[\frac{1}{r}-\frac{f^{\prime }\left( r\right) }{f\left(
r\right) }\right]%
\frac{R^{\prime }\left( r\right) }{R\left( r\right) }-\frac{\left( \ell
-\chi k\right) ^{2}}{r^{2}}  
+M\left( r\right) -k^{2}\nonumber \\
&& 
+f\left( r\right) \left[
E-V\left( r\right) \right] =0,
\end{eqnarray}%
To eliminate the first derivative of $R\left( r\right) $ we may define $%
R\left( r\right) =\sqrt{\frac{f\left( r\right) }{r}}U\left( r\right) $ to
eventually, imply

\begin{eqnarray}  \label{eq:U1}
\frac{U^{\prime \prime }\left( r\right) }{U\left( r\right) }-\frac{\left(
\ell -\chi k\right) ^{2}-1/4}{r^{2}}&+&\frac{1}{4}\left( \frac{f^{\prime
\prime }\left( r\right) }{f\left( r\right) }+\frac{f^{\prime }\left(
r\right) }{r\,f\left( r\right) }\right)-\frac{5}{16}\left( \frac{f^{\prime
}\left( r\right) }{f\left( r\right) }\right) ^{2}\nonumber \\\nonumber \\
&&
-k^{2}+f\left( r\right) %
\left[ E-V\left( r\right) \right] =0,
\end{eqnarray}

It is clear that for constant mass settings $f\left( r\right) =1$, equation (\ref{eq:U1}) collapses into that of (2) in \cite{44} as should be. However, we are interested in the case where $f\left(r\right) \neq const.$ 

At this point, let us consider a power-law type dimensionless scalar multiplier so that%
\begin{equation}  \label{eq:m1}
f\left( r\right) =\lambda r^{\sigma }
\end{equation}%
This would, through \eqref{eq:U1}, imply%
\begin{equation}  \label{eq:U2}
\frac{U^{\prime \prime }\left( r\right) }{U\left( r\right) }-\frac{\left(
\ell -\chi k\right) ^{2}+\frac{\sigma ^{2}}{16}-1/4}{r^{2}}+\lambda
r^{\sigma }\left[ E-V\left( r\right) \right] -k^{2}=0,
\end{equation}%
Equation \eqref{eq:U2} can be solved for different PDM and interaction potential settings. Thus, we treat, in what follows, a particular PDM setting for two cases a free PDM particle and a confined PDM particle by a specific potential. Two illustrative examples are in order.

\subsection{Quasi-free PDM Schr\"{o}dinger particles with space-like screw dislocation}

Here, we consider our quasi-free PDM Schr\"{o}dinger particles (i.e., not subjected to any interaction potential and $V(r) =0$) moving under influence of their own byproducted force field in a space-like screw dislocation. Under such settings, equation \eqref{eq:U2} yields%
\begin{equation}  \label{eq:U3}
\frac{U^{\prime \prime }\left( r\right) }{U\left( r\right) }-\frac{\left(
\ell -\chi k\right) ^{2}+\frac{\sigma ^{2}}{16}-1/4}{r^{2}}+\lambda
r^{\sigma }E-k^{2}=0.
\end{equation}%
Notably, two choices for $\sigma$ are of fundamental interest/nature. %

\subsubsection{A PDM-byproduct Coulomb-like Schr\"{o}dinger model for $\sigma=-1$ and $f(r)=\lambda/r$}  

Let us consider $\sigma =-1 $, which belongs to a Coulombic-like PDM with a scalar multiplier $f\left( r\right) =\lambda /r.$ Hence, the PDM-Schr\"{o}dinger equation \eqref{eq:U3} would read%
\begin{equation}  \label{eq:U4}
\left[ -\frac{d^{2}}{dr^{2}}+\frac{\widetilde{\ell }^{2}-1/4}{r^{2}}-\frac{%
\lambda E}{r}\right] U\left( r\right) =\widetilde{E}U\left( r\right) ,
\end{equation}%
where $\widetilde{\ell }^{2}=\left( \ell -\chi k\right) ^{2}+\frac{1}{16},\ $and $\widetilde{E}=-k^{2}$. Equation \eqref{eq:U4} obviously resembles the two-dimensional radial Coulombic Schr\"{o}dinger equation that has an exact textbook solution. Therefore, the exact eigenvalues are obtained, following the same analysis of [49], as%
\begin{figure}[!ht]
\centering
\includegraphics[width=0.4\textwidth]{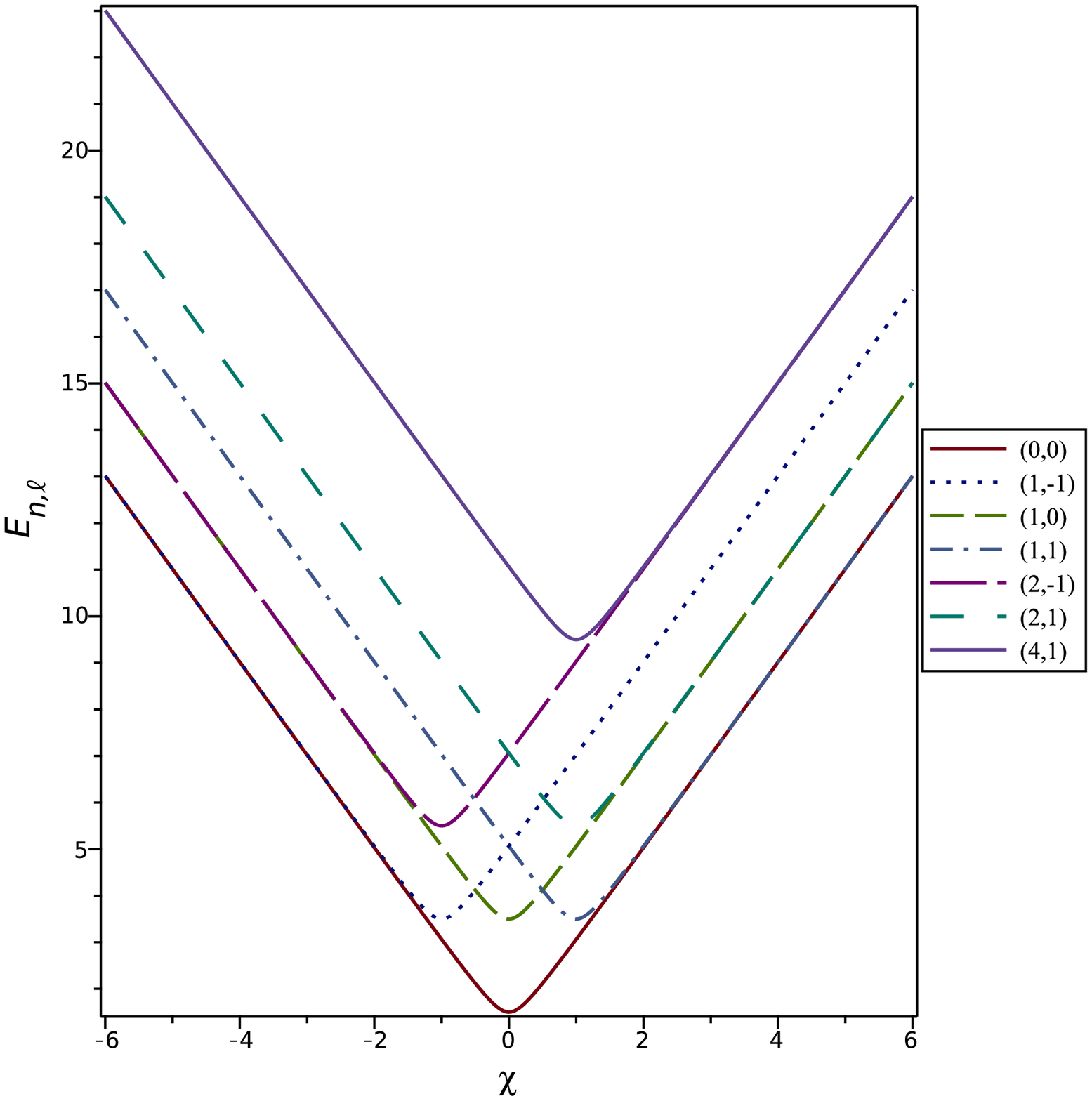}
\caption{The energy levels $\left( n_{r},\ell \right) $ of \eqref{eq:E15} versus the dislocation parameter $\chi$.}
\label{fig:E15}
\end{figure}%
\begin{equation*}  \label{eq:E15}
E_{n_{r},\ell }=\frac{2k}{\lambda }\left[ n_{r}+\left\vert \widetilde{\ell }%
\right\vert +\frac{1}{2}\right]
\end{equation*}to imply%
\begin{equation}
E_{n_{r},\ell }=\frac{2k}{\lambda }\left\{ \left[ \left( \ell -\chi k\right)
^{2}+\frac{1}{16}\right] ^{1/2}+n_{r}+\frac{1}{2}\right\}.
\end{equation}%
Moreover, the exact radial wave functions as%
\begin{equation}  \label{eq:R15}
R\left( r\right) =Nr^{\left\vert \widetilde{\ell }\right\vert -1/2}\ \exp
\left( -kr\right) \ L_{n_{r}}^{2\left\vert \widetilde{\ell }\right\vert
}\left( 2kr\right); \,\, {|\widetilde{\ell }|}\geq \frac{1}{2},
\end{equation}%
where $n_{r}=0,1,2,\cdots$ is the radial quantum number and $%
L_{n_{r}}^{2\left\vert \widetilde{\ell }\right\vert }\left( 2kr\right) $ are the associated Laguerre polynomials.%
To understand the effect of the screw dislocation on the energy levels of the PDM particle, we plot in Fig.1 the energy spectrum of \eqref{eq:E15} versus the screw dislocation parameter $\chi $ related to torsion for different random states (labeled $\left( n_{r},\ell \right) $). We observe that there are multiple energy level crossings for most quantum states mentioned in this figure. For example, the state (1,-1) crosses with the most mentioned states, thus, these states are sharing the same
energy with the state (1,-1) at the crossing points. This would, in turn, indicate so-called occasional degeneracies. On the other hand, there are also infinite degeneracies associated with the term $(\ell -\chi k)^2$ in \eqref{eq:E15} where $(-|\ell|-|\chi|k)^2=(|\ell|+|\chi|k)^2$  for $\ell=\pm |\ell|,\, \chi=\pm|\chi|$, as obvious from Figure 1. In addition to that, there is a notable effect of the torsion parameter $\chi $ that Fig.\ref{fig:E15} as it shows is the rapid increase in the energies as the strength of torsion parameter $|\chi|$ grows up from zero.%

\subsubsection{A PDM byproduct harmonic oscillator-like Schr\"{o}dinger model for $\sigma=2$ and $f(r)=\lambda\,r^2$ }
For $\sigma=2$ and $f(r)=\lambda\,r^2$ %
\begin{equation} \label{eq:U15b}
\left[ -\frac{d^{2}}{dr^{2}}+\frac{\overline{\ell }^{2}-1/4}{r^{2}}-\lambda
Er^{2}\right] U\left( r\right) =\tilde{E}U\left( r\right) 
\end{equation}%
where $\overline{\ell }^{2}=\left( \ell -\chi k\right) ^{2}+\frac{1}{4},$ and $\tilde{E}=-k^{2}.$ This equation\eqref{eq:U15b} is the well-known two-dimensional radial harmonic oscillator problem that has exact solutions. Thus, we follow the same analysis of Gasiorowicz \cite{54} to obtain the energy spectrum as%
\begin{figure}[!h]
\centering
\includegraphics[width=0.4\textwidth]{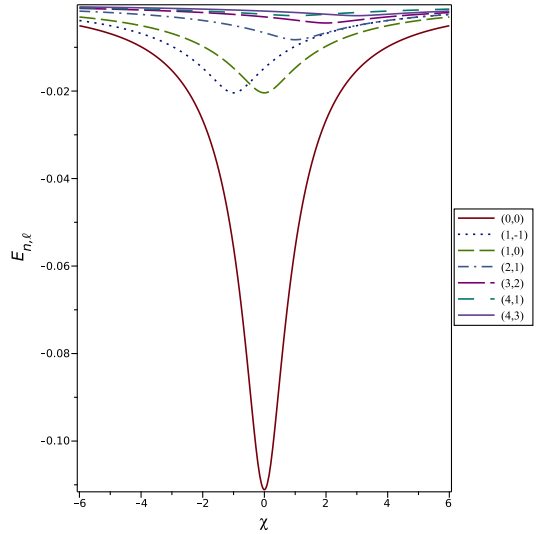}
\caption{The energy levels $\left( n_{r},\ell \right) $ of \eqref{eq:E15b} versus the dislocation parameter $\chi$.}
\label{fig:E15b}
\end{figure}%
\begin{equation}\label{eq:E15b}
E_{n_{r},\ell }=-\frac{1}{\lambda }\left[ \frac{k^{2}}{2\left( \left[ \left(
\ell -\chi k\right) ^{2}+\frac{1}{4}\right] ^{1/2}+2n_r+1\right) }\right]
^{2}
\end{equation}%
and the wave functions read as%
\begin{equation}\label{eq:R15b}
R\left( r\right) =Nr^{-1+\left\vert \overline{\ell }\right\vert }\ \exp
\left( -\frac{\sqrt{-\lambda E_{n_r,\ell}}}{2}r^{2}\right) \ _{1}F_{1}\left(
-n_{r};\left\vert \overline{\ell }\right\vert +1;\sqrt{-\lambda E_{n_r,\ell}}\,
r^{2}\right); \,\, |\overline{\ell }|\geq 1.
\end{equation}%
Fig.\ref{fig:E15b} shows the effect of the PDM setting harmonic type on the energy spectrum, which yields negative values of energy that are opposite to the previous case. However, the energy levels in Fig.\ref{fig:E15b} have similar behavior as in Fig.\ref{fig:E15}, where it can clearly see the energy levels crossing and observed that the minimum energy value when $\chi =0$.

\subsection{PDM Schr\"{o}dinger particles under the influence of space-like screw dislocation and an interaction potential}

For this case, we subject this PDM Schr\"{o}dinger particles to a radial potential of the form%
\begin{equation}  \label{eq:V}
V\left( r\right) =a+b\,r+c\,r^{2},
\end{equation}%
where $a,b,$ and $c$ are the real constants, this kind of scalar potential is constructed from radial potentials such as a linear-type plus a harmonic-type potential, respectively. By substituting this potential into equation \eqref{eq:U2}, we obtain%
\begin{equation}  \label{eq:U5}
\left[ \frac{d^{2}}{dr^{2}}-\frac{\left( \ell -\chi k\right) ^{2}+\frac{%
\sigma ^{2}}{16}-1/4}{r^{2}}+\lambda Er^{\sigma }-\lambda r^{\sigma }\left(
a+br+cr^{2}\right) -k^{2}\right] U\left( r\right) =0.
\end{equation}%
Such a potential is most suited for the one-dimensional PDM-Schr\"{o}dinger form \eqref{eq:U2} and is anticipated to be exactly solvable for a sample of
PDM settings. Therefore, we treat, in what follows, a particular radial PDM settings, so that its exact solutions are inferred from some models that are
known to be exactly solvable.%
Taking the case where $\sigma =-2$ leads to the PDM setting $f\left( r\right) =\lambda r^{-2}.$ Thus a Coulomb-like Schr\"{o}dinger model is obtained as%
\begin{equation}  \label{eq:U6}
\left[ -\frac{d^{2}}{dr^{2}}+\frac{\mathcal{L} ^{2}-1/4}{r^{2}}-\frac{
\lambda b}{r}\right] U\left( r\right) =\bar{E}U\left( r\right) ,
\end{equation}%
where $\mathcal{L} ^{2}=\left( \ell -\chi k\right) ^{2}-\lambda \left(
E-a\right) +\frac{1}{4},$ and $\bar{E}=\lambda c+k^{2}.$ Moreover, we have again a similar two-dimensional radial Schr\"{o}dinger equation of Coulombic nature that admits exact eigenvalues that yields%
\begin{figure}[!ht]
\subfloat[\label{subfig:E20a}]{    
 \includegraphics[width=0.3\textwidth]{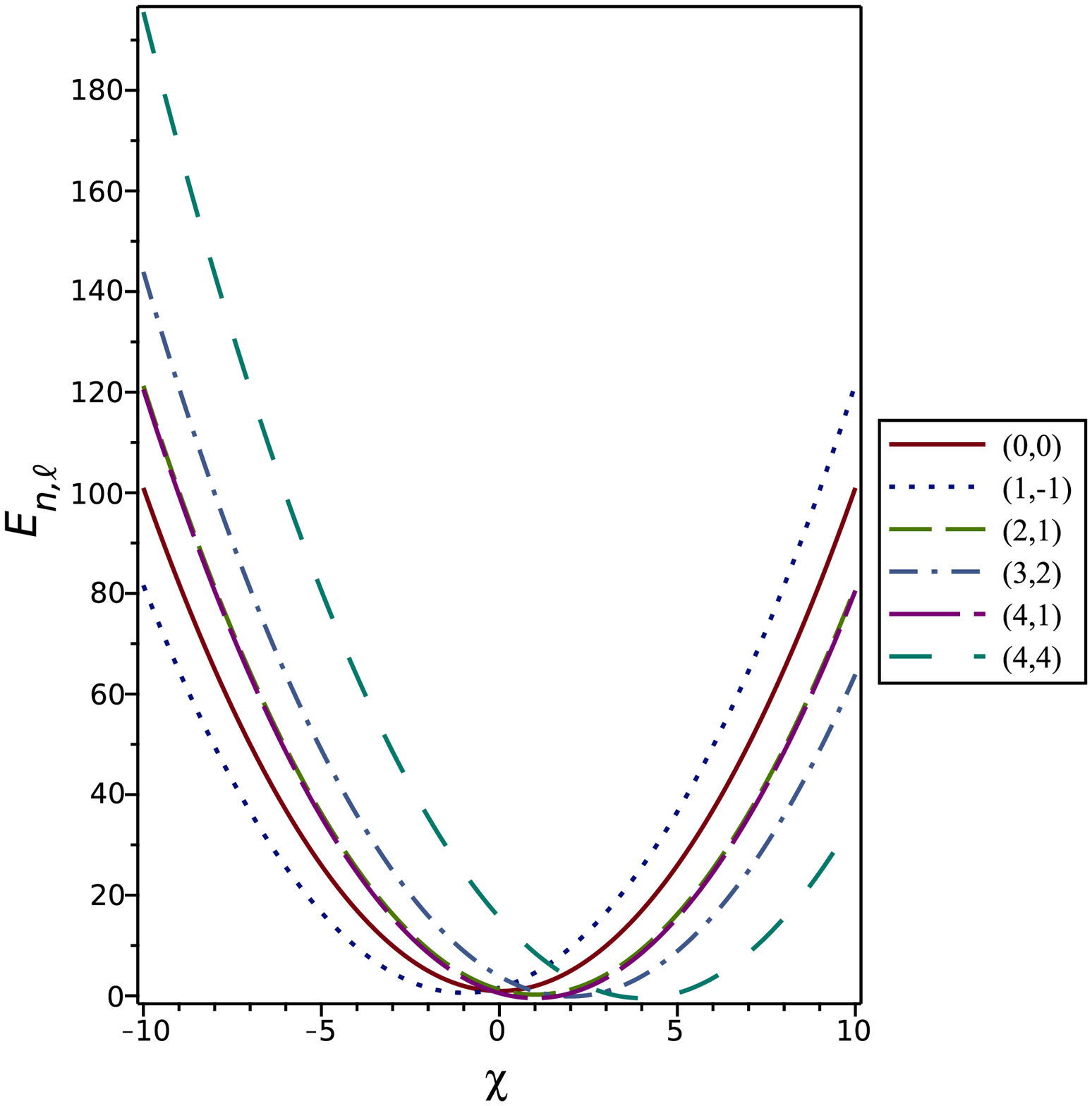} } 
\subfloat[\label{subfig:E20b}]{\includegraphics[width=0.3\textwidth]{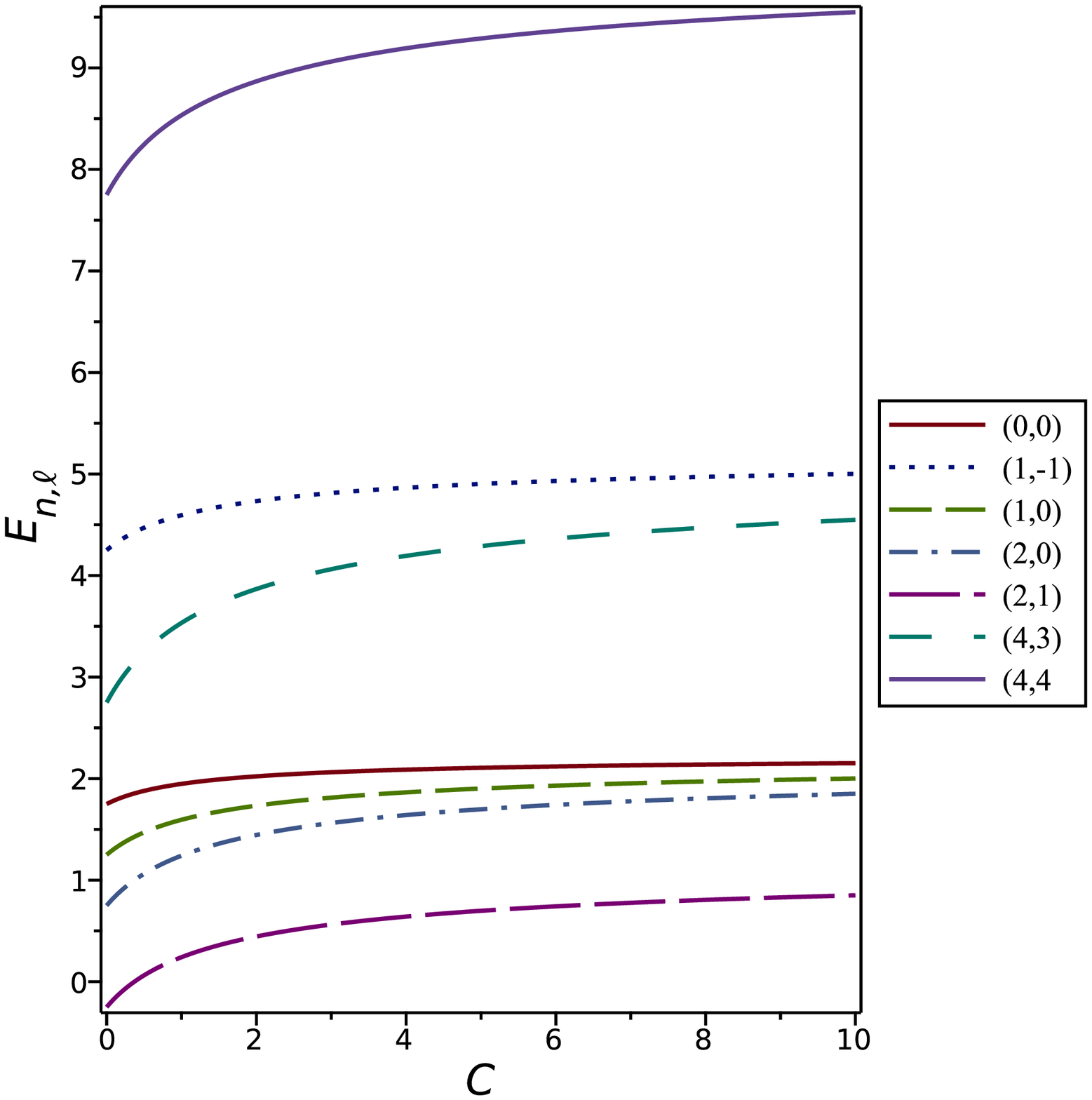}}
\subfloat[\label{subfig:E20c}]{\includegraphics[width=0.3\textwidth]{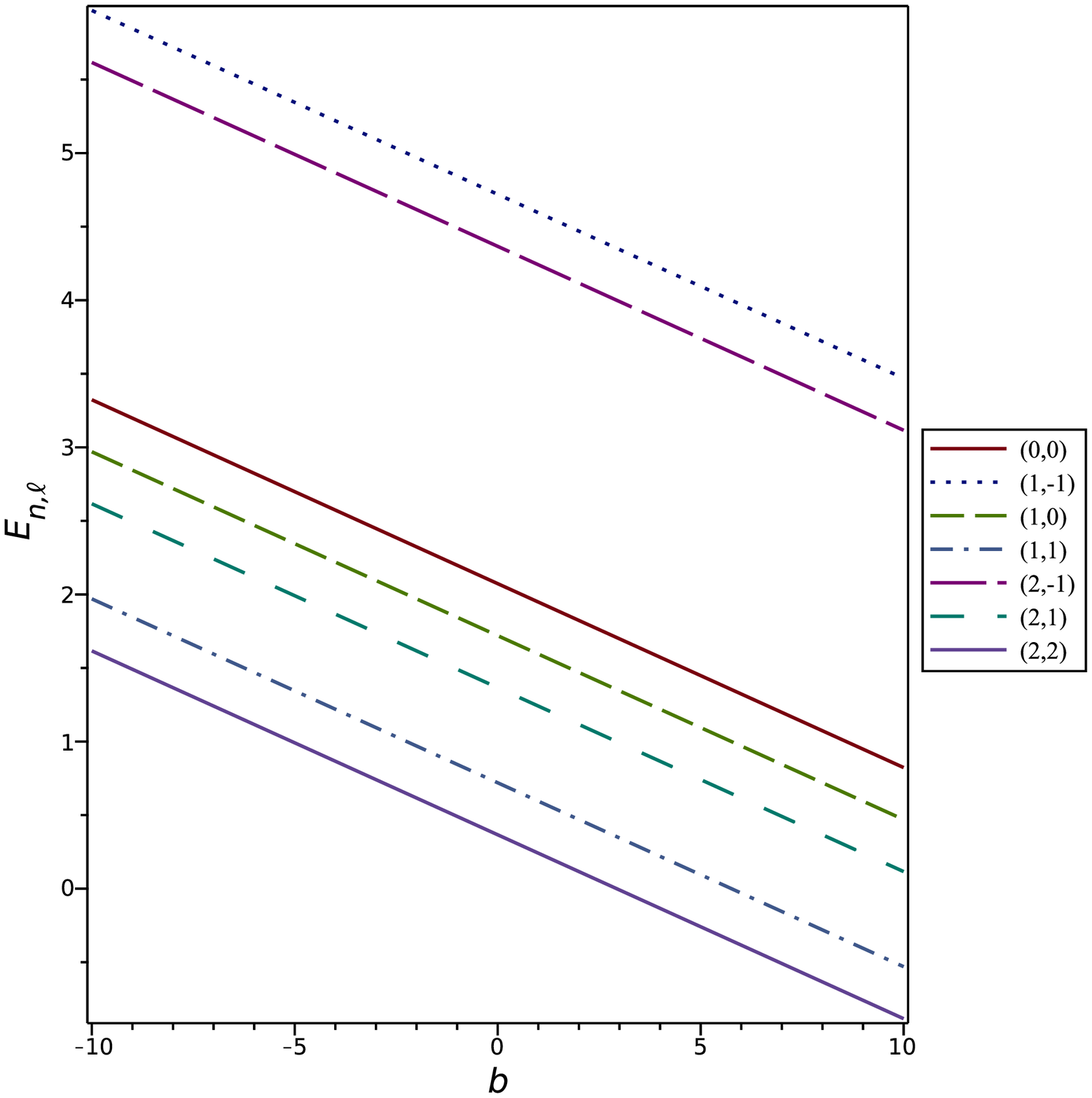} }
\caption{The energy levels $\left( n_{r},\ell \right) $ of \eqref{eq:E20} versus (a) the dislocation parameter $\chi$, (b) the parameter of the harmonic potential part $c
$, and (c) the parameter of the linear potential part $b$ .}
\label{fig:E20}
\end{figure}
\begin{equation}  \label{eq:E20}
E_{n_{r},\ell }=\frac{1}{\lambda }\left\{ \left( \ell -\chi k\right)
^{2}+\lambda a+\frac{1}{4}-\frac{\left[ \lambda b+\sqrt{\lambda c+k^{2}}%
\left( 2n_{r}+1\right) \right] }{4\left( \lambda c+k^{2}\right) }\right\}
\end{equation}%
and radial eigenfunctions read as%
\begin{equation}
R\left( r\right) =Nr^{-1+\left\vert \mathcal{L} \right\vert }\ \exp
\left( -\sqrt{\lambda c+k^{2}}r\right) \ L_{n_{r}}^{2\left\vert \mathcal{L} \right\vert }\left( 2\sqrt{\lambda c+k^{2}}r\right);\, |\mathcal{L}|\geq 1.   \label{eq:R21}
\end{equation}%
Because of the presence of the effect of potential force field $V(r)$ on the system, it generally leads to a shift in energy levels, but each part of this potential has an effect opposite to the other, where the first part gradually increases the energy values and the second one decreases them as shown in the two 
Figures 3b and 3c. In this case, there are also multiple energy level crossings ( i.e., occasional degeneracies), as shown
in Fig.3. However, the shifts in the energy levels that occurred due o the combined effects of the screw dislocation and the scalar potential produce a different pattern of the energy levels crossings. Additionally,  there are infinite degeneracies between some of the levels for all values of $\left\vert \chi \right\vert$. ( between the state (2,1) and (4,1) as an example).

\section{Charged PDM-Schr\"{o}dinger particles in space-like dislocation, magnetic and Aharonov-Bohm flux fields}

In this section, we analyze the moving of a PDM charged particle in an elastic medium with a screw dislocation subjected to magnetic and Aharonov-Bohm flux (AB) fields. The PDM Hamiltonian corresponds to a positively charged particle in the presence of a screw dislocation in Eq (\ref{eq:SC}) and vector potential in the background given by the metric $g^{ij}$, is written as%
\begin{equation}  \label{eq:H2}
\hat{H}=\frac{1}{\sqrt{g}}\left( \frac{\hat{p}_{i}-q\,A_{i}}{\sqrt{f(r)}}\right) \sqrt{g}\,g^{ij}\left( \frac{\hat{p}_{j}-q\,A_{j}}{\sqrt{%
f(r) }}\right)
\end{equation}%
By use of the minimal coupling of the PDM-momentum and the vector potential $\hat{\mathbf{p}}\rightarrow \hat{\mathbf{p}}-q\mathbf{A}$. Where $q$ is the electrical charge of the particle, $\hat{\mathbf{p}}$ is again
the PDM- momentum operator in \eqref{eq:P}, and $A_{i}$ is a covariant component of the electromagnetic four-vector potential $A_{\mu }=\left(
A_{0},\mathbf{A}\right) $ with the vector potential:%
\begin{equation}  \label{eq:A1}
\mathbf{A} =\left( 0,\frac{B_{\circ }%
}{2}r+\frac{\Phi _{AB}}{2\pi r},0\right) ,
\end{equation}%
that includes Aharonov-Bohm flux field $\Phi _{AB}$. Consequently, equation \eqref{eq:H2} would result%
\begin{eqnarray}  \label{eq:m2}
&&-\frac{1}{m\left( r\right) }\left[ \partial _{r}^{2}+\left[\frac{1}{r}-\frac{f^{\prime }\left( r\right) }{f\left(
r\right) }\right]\partial
_{r}+\frac{1}{r^{2}}\left( \partial _{\phi }-\chi \partial _{z}\right)
^{2}  \right. \nonumber \\ \nonumber \\ 
&& \left.
+\partial _{z}^{2}-\frac{2iqA_{\phi }}{r^{2}}\left( \partial _{\phi
}-\chi \partial _{z}\right) 
-\frac{q^{2}A_{\phi }^{2}}{r^{2}}+M\left(
r\right) \right] \psi +V\left( r\right) \psi =E\psi
\end{eqnarray}%
Next, we use the same way as the previous section to determine the non-relativistic PDM-Schr\"{o}dinger equation for this case in cylindrical
coordinates and use the wavefunction $\psi \left( r,\phi ,z\right)
=e^{i\ell \phi }e^{ikz}R\left( r\right)$ to obtain%
\begin{eqnarray}
\frac{R^{\prime \prime }\left( r\right) }{R\left( r\right) }+\left[\frac{1}{r}-\frac{f^{\prime }\left( r\right) }{f\left(
r\right) }\right]%
\frac{R^{\prime }\left( r\right) }{R\left( r\right) }-\frac{\left( \ell
-\chi k\right) ^{2}}{r^{2}}+qB_{\circ }\left( \ell -\chi k\right) +\frac{%
q\Phi \left( \ell -\chi k\right) }{\pi r^{2}}  \nonumber \\  \nonumber \\ 
-\frac{q^{2}\Phi ^{2}}{4\pi ^{2}r^{2}}-\frac{q^{2}B_{\circ }^{2}r^{2}}{4}-%
\frac{q^{2}\Phi B_{\circ }}{2\pi }+M\left( r\right) -k_{z}^{2}+f\left(
r\right) \left[ E-V\left( r\right) \right] =0,  \label{R2}
\end{eqnarray}%
Where the interaction of the charged PDM particle with the uniform magnetic field $\mathbf{B}= \mathbf{\nabla} \times \mathbf{A}=B_{\circ }\hat{z}$ yields a discrete spectrum of energy, which is called the Landau levels \cite{54}. 

Using the same transformation recipe $R\left( r\right) =\sqrt{\frac{f\left(r\right) }{r}}U\left( r\right) $ to eliminate the first derivative of $%
R\left( r\right) $ equation \eqref{R2} reads

\begin{gather}
\frac{U^{\prime \prime }\left( r\right) }{U\left( r\right) }-\frac{\left[
\left( \ell -\chi k-\frac{q\Phi }{2\pi }\right) ^{2}-\frac{1}{4}\right] }{%
r^{2}}+\frac{1}{4}\left( \frac{f^{\prime \prime }\left( r\right) }{f\left(
r\right) }+\frac{f^{\prime }\left( r\right) }{r\,f\left( r\right) }\right) -%
\frac{5}{16}\left( \frac{f^{\prime }\left( r\right) }{f\left( r\right) }%
\right) ^{2} \nonumber \\ \nonumber \\ 
+qB_{\circ }\left( \ell -\chi k-\frac{q\Phi }{2\pi }\right) -\frac{%
q^{2}B_{\circ }^{2}r^{2}}{4}-k_{z}^{2}+f\left( r\right) \left[ E-V\left(
r\right) \right] =0,  \label{U7}
\end{gather}%
As in the previous section, In the following illustrative examples, we focus on the analysis of the effect of the screw dislocation on the PDM charged particle in
the presence of magnetic and Aharonov-Bohm flux fields, by using the same power-law PDM setting in \eqref{eq:m1} for the two cases: an almost quasi-free and confined PDM charged particles by the scalar potential \eqref{eq:V}.

\subsection{An almost quasi-free PDM-Schr\"{o}dinger charged particle $V\left( r\right) =0:$}

Let us consider an almost quasi-free PDM positively charged particle with $f\left(r\right) =\lambda r^{\sigma }$ moving in the vector potential \eqref{eq:A1}. Thus, equation \eqref{U7} reads%
\begin{eqnarray}  \label{eq:U8}
\left[ \frac{d^{2}}{dr^{2}}-\frac{\left( \ell -\chi k-\frac{q\Phi }{2\pi }%
\right) ^{2}+\frac{\sigma ^{2}}{16}-\frac{1}{4}}{r^{2}}-\frac{q^{2}B_{\circ
}^{2}r^{2}}{4}+\lambda Er^{\sigma } \right.\nonumber \\ \nonumber \\  \left.
+\left[ qB_{\circ }\left( \ell -\chi k-%
\frac{q\Phi }{2\pi }\right) -k^{2}\right] \right] U\left( r\right) =0.
\end{eqnarray}%
For $\sigma =-2$ , one would obtain

\begin{equation}  \label{eq:U9}
\left[ -\frac{d^{2}}{dr^{2}}+\frac{\tilde{m}^{2}-\frac{1}{4}}{r^{2}}+%
\frac{q^{2}B_{\circ }^{2}r^{2}}{4}\right] U\left( r\right) =\tilde{\lambda}U\left(
r\right),
\end{equation}%
where $\omega=\frac{q\,B_{\circ}}{2}\geq\,0$, $\tilde{m}^{2}=\left( \ell -\chi k-\frac{q\Phi }{2\pi }\right)^{2}-\lambda E+\frac{1}{4}$ and $\ \widetilde{\lambda }=\left[ qB_{\circ}\left( \ell -\chi k-\frac{q\Phi }{2\pi }\right) -k^{2}\right]$.  Equation\eqref{eq:U9} is, again, in the form of the well-known two-dimensional radial harmonic oscillator problem and admits the exact solution of the form%
\begin{equation}
\tilde{\lambda}=2\,\omega(2\,n_r+|\tilde{m}|+1) \label{U9-1}
\end{equation}%
\begin{figure}[!ht]
\subfloat[\label{subfig:E29a}]{      
 \includegraphics[width=0.4\textwidth]{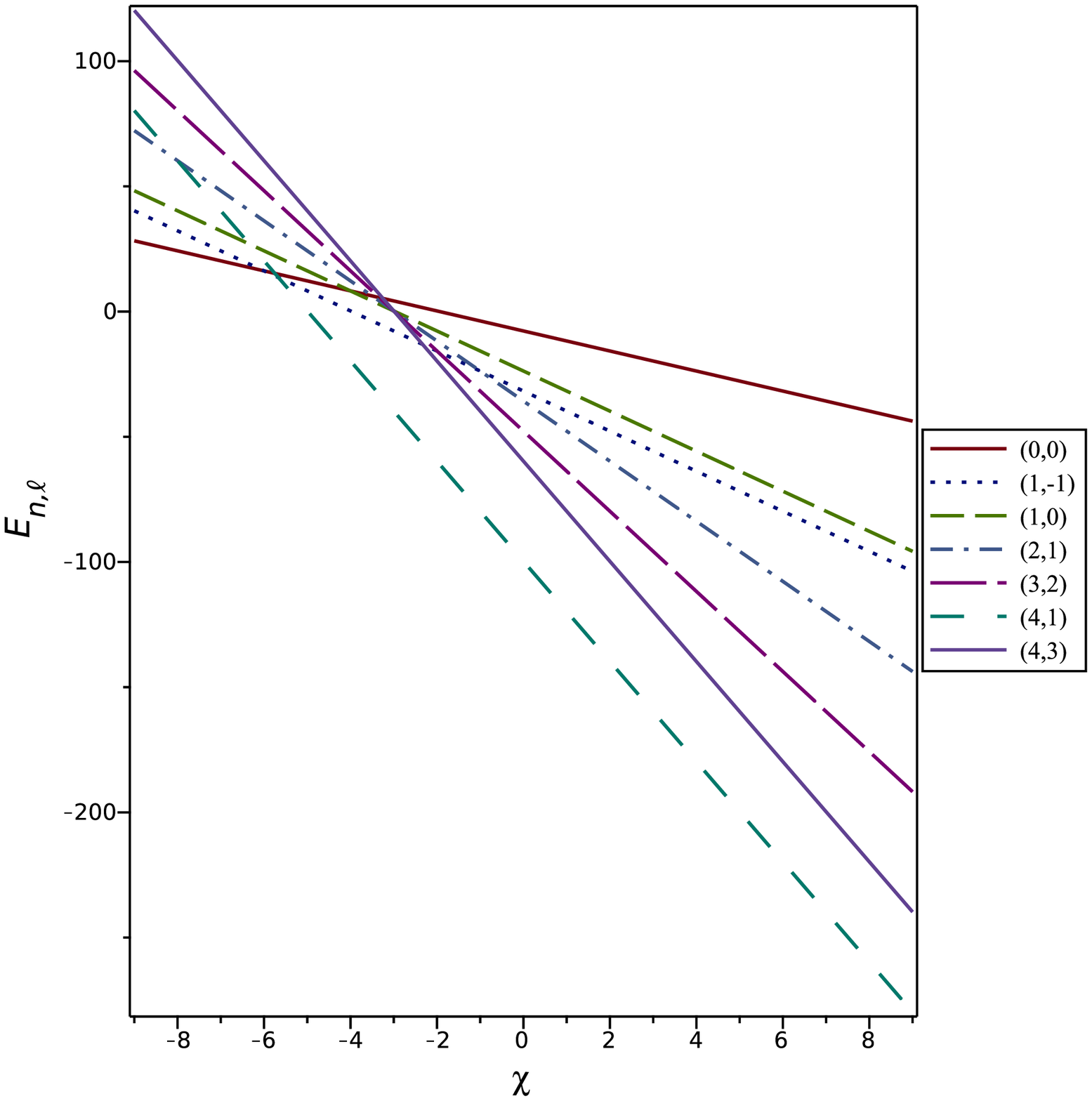} } 
\subfloat[\label{subfig:E29b}]{\includegraphics[width=0.4\textwidth]{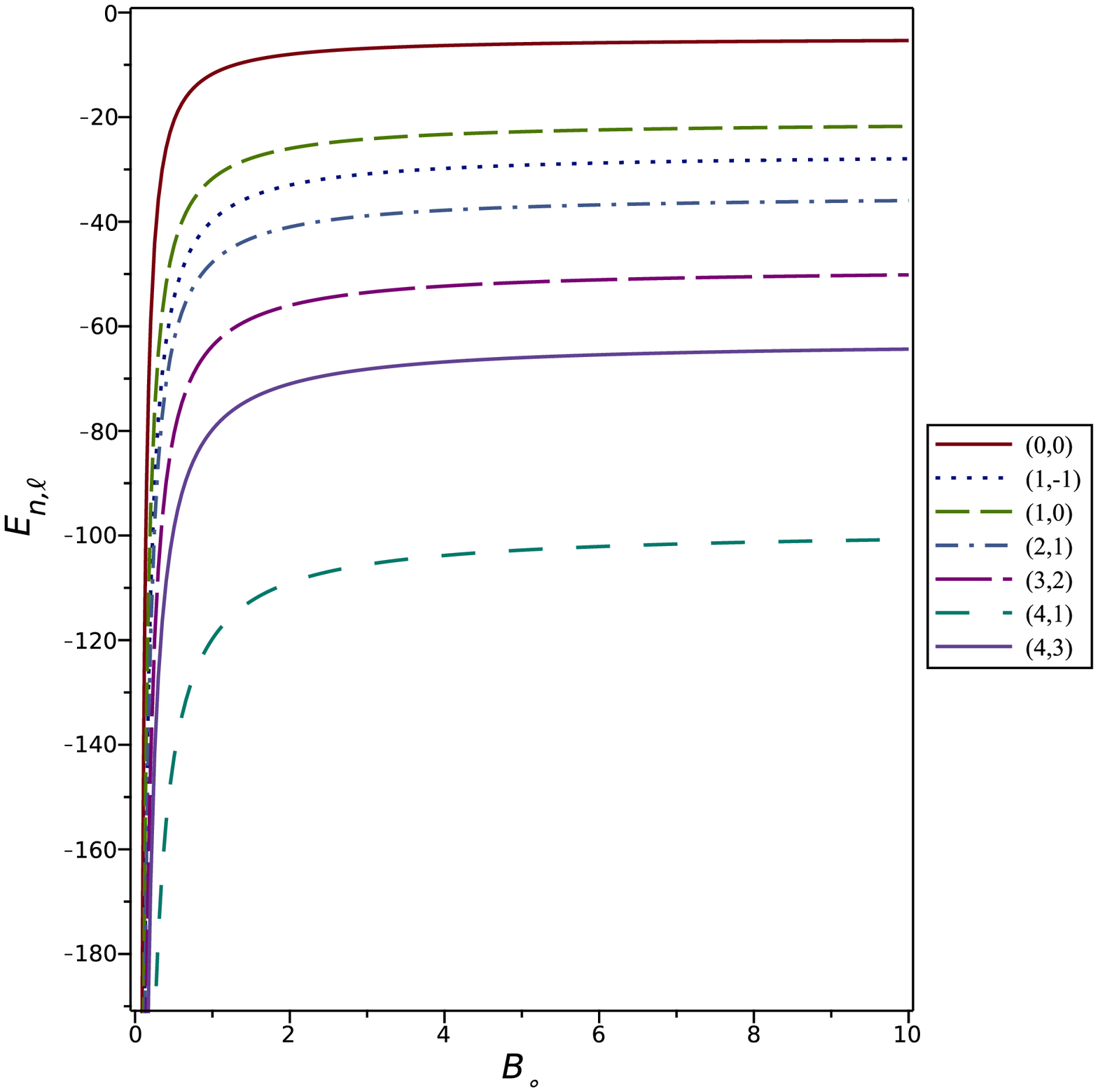} }
\caption{The energy levels $\left( n_{r},\ell \right) $ of \eqref{eq:E29} versus (a) the dislocation parameter $\chi$, and (b) the magnetic field parameter $B_{\circ }$ .}
\label{fig:E29}
\end{figure}%
to yield the eigenvalues%
\begin{equation}
E_{n_{r},\ell }=\frac{1}{\lambda }\left\{ \left( \ell -\chi k-\frac{q\Phi }{%
2\pi }\right) ^{2}-\left[ \left( \ell -\chi k-\frac{q\Phi }{2\pi }\right)
-k^{2}/qB_{\circ }-2n_{r}-1\right] ^{2}+\frac{1}{4}\right\}   \label{eq:E29}
\end{equation}%
and the corresponding eigenfunctions as%
\begin{equation}  \label{eq:R30}
R\left( r\right) =Nr^{-1+\left\vert \tilde{m}\right\vert }\ \exp \left( -\frac{qB_{\circ }}{4}r^{2}\right) \ _{1}F_{1}\left( -n_{r};\left\vert 
\tilde{m}\right\vert +1;\frac{qB_{\circ }}{2}r^{2}\right); \,\, |\tilde{m}|.
\end{equation}%

In Fig.4, we observe that the influence of the screw dislocation with the presence of the magnetic and Aharonov-Bohm fields breaks the infinite
degeneracies of the energy levels, where Fig.4a, shows only multiple energy levels crossing for some states. We observe, in Fig.4a, that the energy spectrum has negative values in the positive $\chi $ region and positive values in the negative $\chi $ region. Furthermore, Fig.4b, shows that the magnetic field $B_{\circ }$ affects the energy spectrum by decreasing the values of the energies and mostly becoming negative.

\subsection{PDM charged particle under a combined effect of screw dislocation and interaction potential}

Now, let us consider our PDM particle, with the same PDM setting \eqref{eq:m1}, moving under the influence of an interaction potential \eqref{eq:V}. Hence, one may rewrite equation \eqref{U7} as%
\begin{eqnarray}  \label{eq:U10}
\left[ \frac{d^{2}}{dr^{2}}-\frac{\left( \ell -\chi k-\frac{q\Phi }{2\pi }%
\right) ^{2}+\frac{\sigma ^{2}}{16}-\frac{1}{4}}{r^{2}}-\frac{q^{2}B_{\circ
}^{2}r^{2}}{4} +\lambda E\,r^{\sigma} \right. \nonumber \\  \nonumber \\ \left.
-\lambda r^{\sigma}\left(
a+br+cr^{2}\right) +\left[ qB_{\circ }\left( \ell -\chi k-\frac{q\Phi }{2\pi 
}\right) -k^{2}\right] \right] U\left( r\right) =0.
\end{eqnarray}%
Choosing $ \sigma =-2 $  in \eqref{eq:U10} would yield%
\begin{equation}  \label{eq:U11}
\left[ \frac{d^{2}}{dr^{2}}-\frac{(\mathcal{\tilde{L}}^2-\frac{1}{4})}{r^{2}}-\omega^{2}r^{2}-\frac{\tilde{b}}{r}+\tilde{\Lambda}\right] U\left( r\right)
=0,
\end{equation}%
where  $\mathcal{L}^2=\tilde{m}^2+\tilde{a}$, $\tilde{\Lambda}=\tilde{\lambda}-\lambda c$, $\tilde{a}=\lambda a$, $\tilde{b}=\lambda b$. Which with the substitution of%
\begin{equation}
U(r)=r^{|\mathcal{\tilde{L}}|+1/2}\, exp(-\frac{\omega\,r^2}{2})\, H(r), \label{Heun0}
\end{equation}%
would result in%
\begin{equation}
rH''(r)+[(2|\mathcal{\tilde{L}}|+1)-2\omega r^2]\,H'(r)-[\zeta \,r+\tilde{b}]H(r)=0. \label{Heun1}
\end{equation}%
Which in turn, upon the substitution of%
\begin{equation}
H(r)=\sum\limits_{j=0}^{\infty }C_{j}\,r^{j}, \label{Heun2}
\end{equation}%
would yield $C_0=1,\, C_1=\frac{\tilde{b}}{(1+2|\mathcal{\tilde{L}}|)}$ and%
\begin{equation}
C_{j+2}[(j+2)(j+2|\mathcal{\tilde{L}}|+2]=\tilde{b}\, C_{j+1}+(2\omega\,j+\zeta)\,C_j;\,\, j=-1,0,1,2,\cdots, \, C_{-1}=0. \label{Heun3}
\end{equation}%
To allow finiteness and square integrability of the wave function, we truncate the biconfluent Heun series into a polynomial of degree $n$ by requiring that $\forall\,{C_j}'s=0$  for $\forall\,j>n$. This would suggest that $C_{n+1}=0=C_{n+2}$ and $C_n (2\omega\,n+\zeta)=0\Rightarrow  \zeta=-2\omega n$. Consequently,%
\begin{equation} \label{Heun4}
\tilde{\Lambda}=2\omega(n+|\mathcal{\tilde{L}}|+1).
\end{equation}%
This result should retrieve that of (\ref{U9-1}) when $\tilde{a}=0=\tilde{b}=c$. In this case, $\mathcal{\tilde{L}}^2=\tilde{m}^2$ and $\tilde{\Lambda}=\tilde{\lambda}$ to imply that the truncation order $n$ is correlated with the radial quantum number $n_r=0,1,2,\cdots$ through the correlation $n=2n_r$  (c.f., e.g.,  \cite{50,51,52,55,56}). Under such settings, one would cast%
\begin{equation} \label{Heun5}
\tilde{\Lambda}=2\omega(2\,n_r+|\mathcal{\tilde{L}}|+1),\,\, H(r)=\sum\limits_{j=0}^{n=2n_r }C_{j}\,r^{j}.
\end{equation}%
Using (\ref{eq:U11}) and (\ref{Heun5}) would give the energy levels%
\begin{eqnarray}  \label{eq:E38}
E_{n_{r},\ell }=\frac{1}{\lambda }\left\{ \left( \ell -\chi k-\frac{q\Phi }{2\pi }\right) ^{2}-\left[ \left( \ell -\chi k-\frac{q\Phi }{2\pi }\right)  \right.\right.  \nonumber \\
 \left. \left.
-\left( k^{2}+\lambda c\right) /qB_{\circ }-2n_{r}-1\right] ^{2}+\lambda a+
\frac{1}{4}\right\},
\end{eqnarray}%
and the radial eigenfunctions%
\begin{equation}  \label{eq:R39}
U(r)=r^{|\mathcal{\tilde{L}}|+1/2}\, exp(-\frac{\omega\,r^2}{2})\, H_{B}\left(2|\mathcal{\tilde{L}}|,0,\frac{\tilde{\Lambda}}{\omega},\frac{2\tilde{b}}{\sqrt{\omega}},\sqrt{\omega}\,r\right).
\end{equation}%
At this point, one should notice that, when $b=0=\tilde{b}$ the biconfluent Heun polynomial of degree $n=2n_r$ reduces into $H_{B}\left(2|\mathcal{\tilde{L}}|,0,\frac{\tilde{\Lambda}}{\omega},0,\sqrt{\omega}\,r\right)\Rightarrow\, _{1}F_{1}\left(2|\mathcal{\tilde{L}}|,0,\frac{\tilde{\Lambda}}{\omega},0,\sqrt{\omega}\,r\right)$ that immediately results (\ref{Heun5}) and consequently the harmonic oscillator eigenvalues are retrieved \cite{50,51,52}. Moreover, the three terms recurrence relation (\ref{Heun3}) would imply%
\begin{equation} \label{Heun6}
C_2=\frac{\tilde{b}^2}{2(2+2|\mathcal{\tilde{L}}|)(1+2|\mathcal{\tilde{L}}|)},\, C_3=\frac{\tilde{b}\,C_2+(2\omega+\zeta)\,C_1}{3(3+2|\mathcal{\tilde{L}}|)}, \cdots ,
\end{equation}%
and so on. Then the coefficients of the polynomial in (\ref{Heun5}) are determined in the process.

Yet, it has been a tradition in the literature that (\ref{Heun3}) is also expressible in the form of%
\begin{equation}
C_{j+1}[(j+1)(j+2|\mathcal{\tilde{L}}|+1]=\tilde{b}\, C_{j}+(2\omega\,(j-1)+2\omega n)\,C_{j-1};\,\, 0\leq j \leq n, \, C_{-1}=0, \label{Heun7}
\end{equation}%
\begin{figure}[!ht]
\subfloat[\label{subfig:E38a}]{      
 \includegraphics[width=0.3\textwidth]{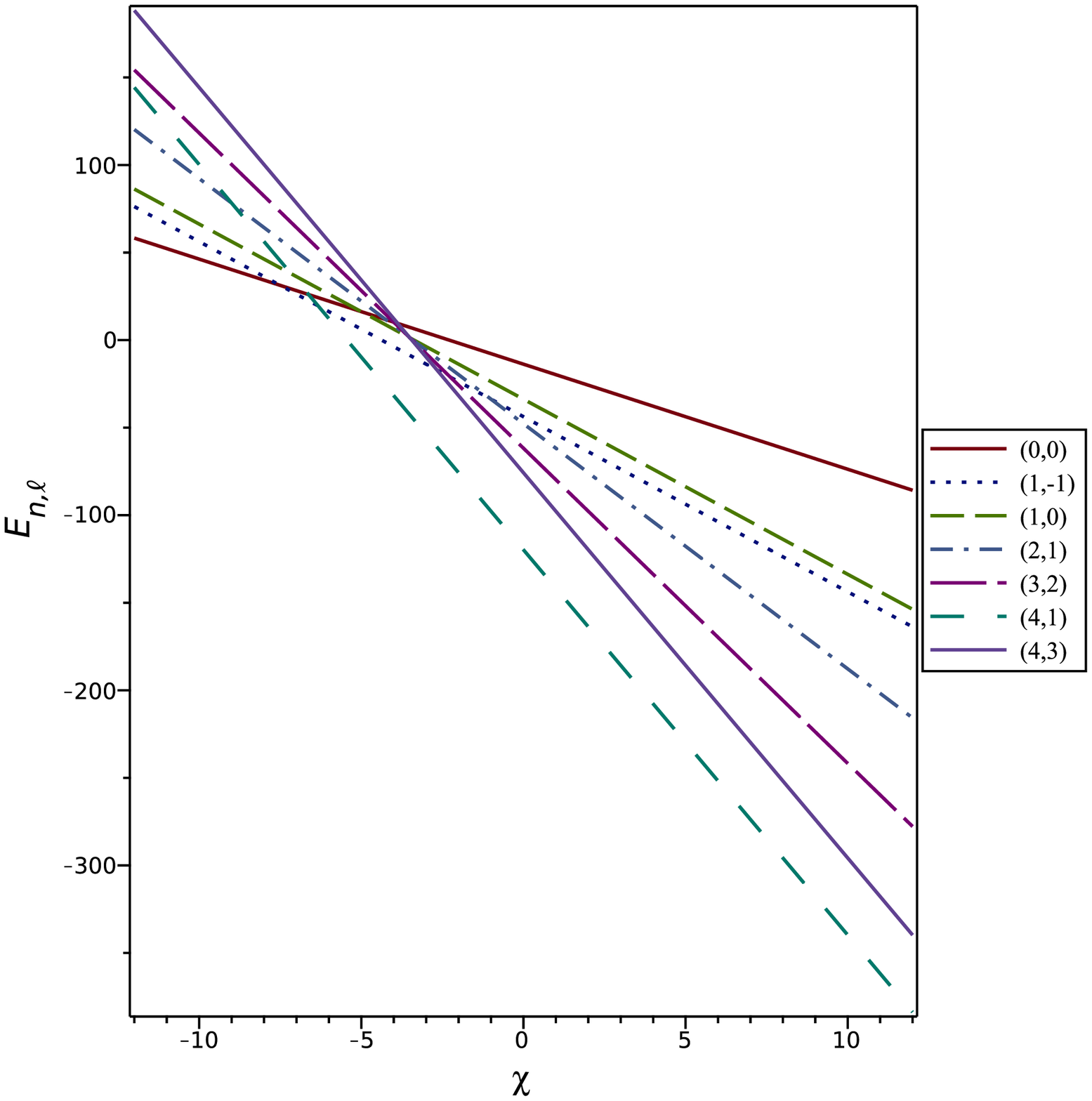}} 
\subfloat[\label{subfig:E38b}]{\includegraphics[width=0.3\textwidth]{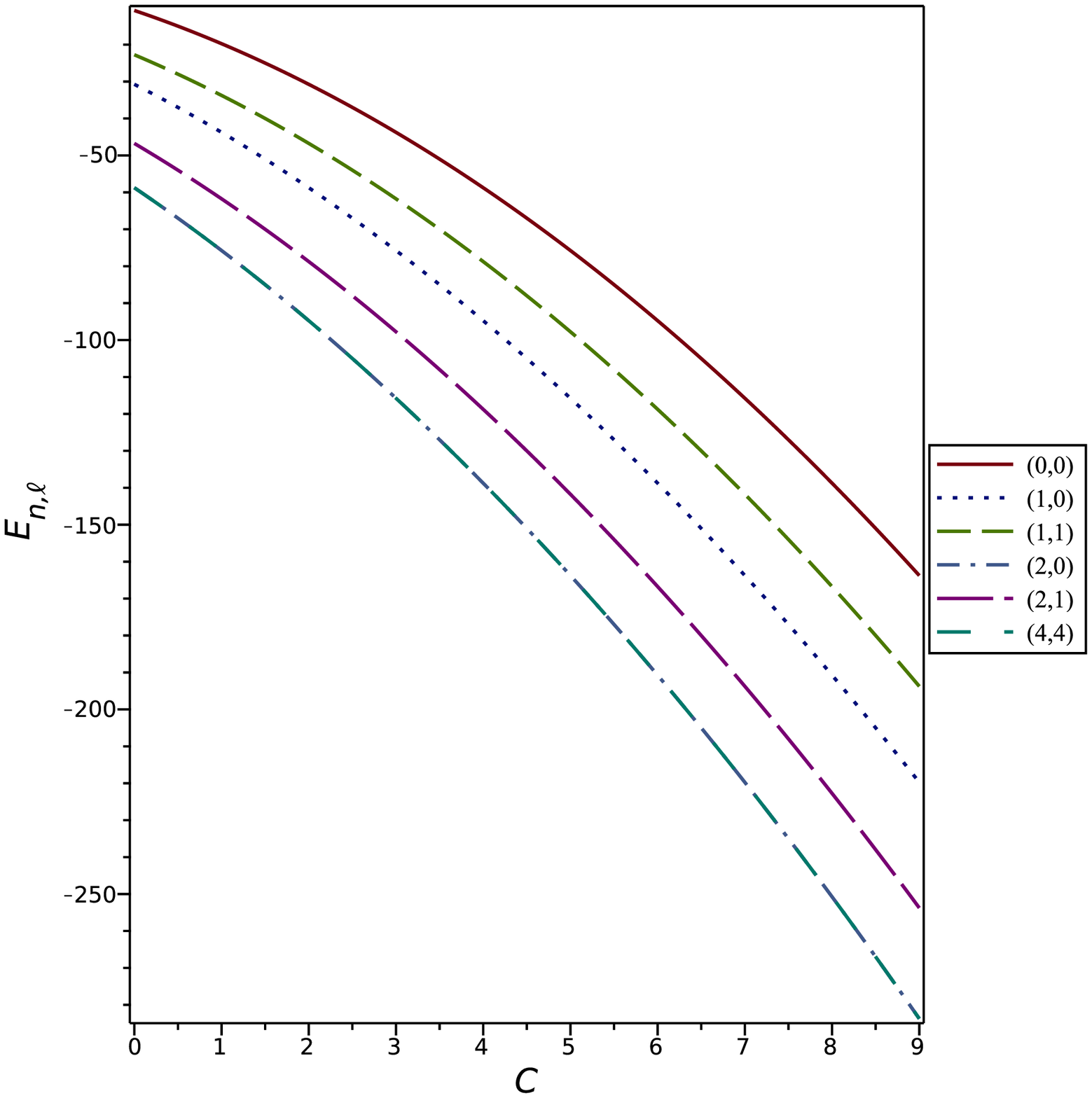} }  
\subfloat[\label{subfig:E38c}]{ \includegraphics[width=0.3\textwidth]{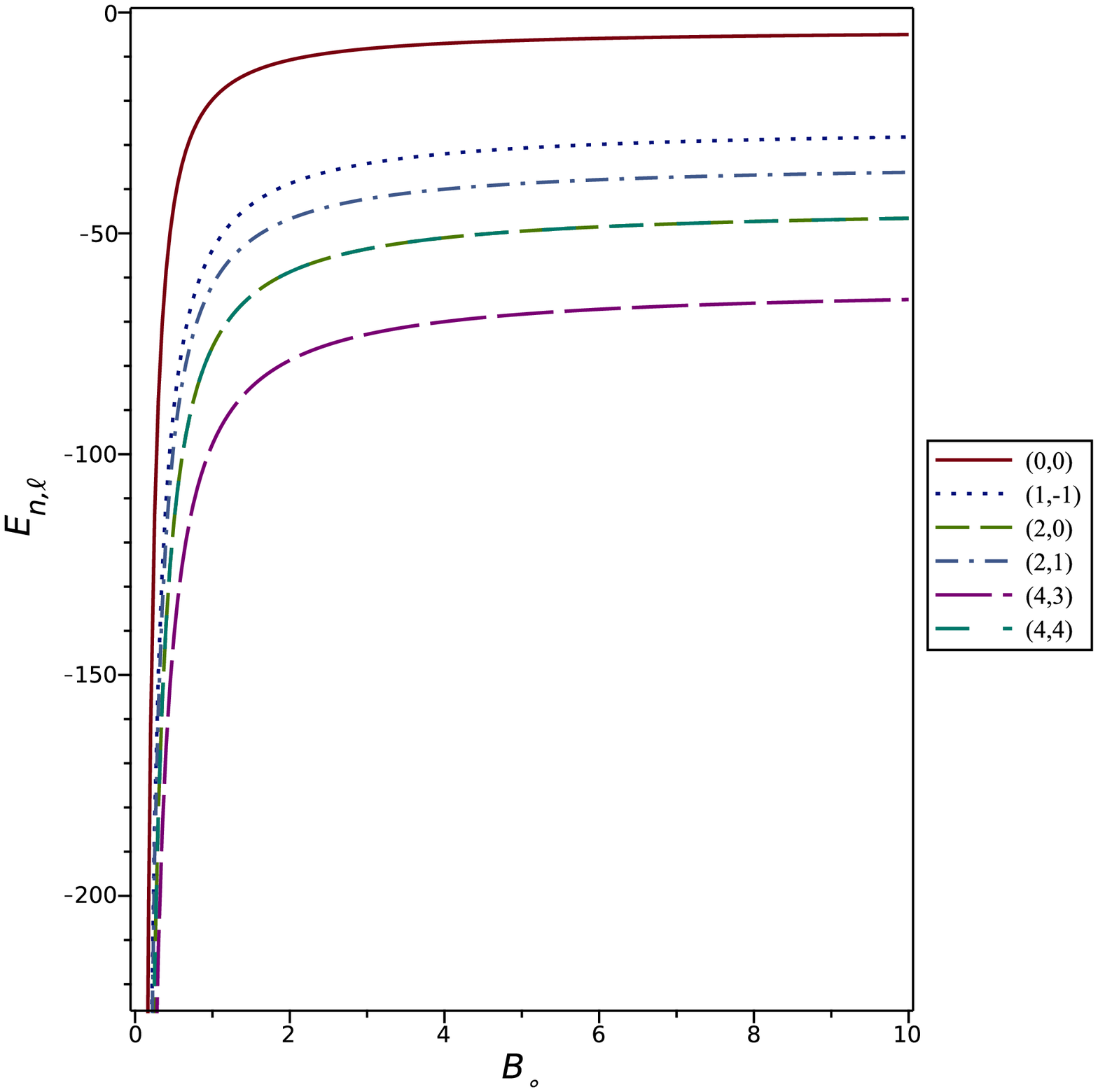}}
\caption{The energy levels $\left( n_{r},\ell \right) $ of \eqref{eq:E38} versus (a) the dislocation parameter $\chi$, (b) the parameter of the harmonic potential part $c
$, and (c) the magnetic field parameter $B_{\circ }$ .}
\label{fig:E38}
\end{figure}%
and reinforcing \textit{again} the condition $C_{n+1}=0$ for $\forall j>n$. The claim is that this would introduce yet another quantization recipe or interrelate the parameters involved (here it would interrelate $\tilde{b}$ with $\omega$). Such assumptions have to be tested in what follows. For $j=n$ and $C_{n+1}=0$ equation (\ref{Heun7}) implies%
\begin{equation}\label{Heun8}
\tilde{b}\, C_n =2\,\omega\, C_{n-1}.
\end{equation}%
Which, in turn, gives%
\begin{equation}\label{Heun9}
\tilde{b}\,C_0=2\omega\,C_{-1}\Rightarrow \tilde{b}=0 \,\, for \,\,n=0,
\end{equation}%
\begin{equation}\label{Heun10}
\tilde{b}\,C_1=2\omega\,C_{0}\Rightarrow \tilde{b}^2=2\,\omega\,(2|\mathcal{\tilde{L}}|+1)\, \,for \,\, n=1,
\end{equation}%
and so on and so forth. Notable, whilst (\ref{Heun9}) kills the Coulombic like term (i.e., $b=0=\tilde{b}$), relation (Heun10) yields that $(2|\tilde{\mathcal{L}}|+1)=0\Rightarrow |\tilde{\mathcal{L}}|=-1/2$ (of course if we do not want a quantum collapse, by the assumption that $\omega=0$, to take place). Obviously, neither mathematically nor physically, $|\tilde{\mathcal{L}}|=-1/2$ can be accepted. One should, therefore, accept what is physically and mathematically safely offered by the biconfluent Heun polynomials related to (\ref{Heun4}), and classify the reinforced-again condition as a mathematically and/or physically redundant one. This is a common practice we always follow in quantum mechanics. Yet, Mustafa \cite{52} has very recently provided brute-force evidence that reinforcing the condition $C_{n+1}=0$   would yield results that are neither mathematically nor physically acceptable (see the detailed analysis on such a model in the appendix of \cite{52}). Of course, $C_{n+1}$ is a polynomial of degree $n+1$ is mathematically correct and interesting for the biconfluent Heun differential equation, but not for quantum mechanics. %

The combined effect of the magnetic field, AB flux field, and scalar potential in the existence of the screw dislocation yield a shift in the energy levels and decreases the energy values, although Figures 5a and 5b, have a similar pattern to Figures 4a and 4b. 
 In particular,  the linear part of the potential has no effect on the eigenvalues \eqref{eq:E38}, on the other hand, the harmonic part has a notable effect on the energy levels where it causes again the infinite degeneracies between some states (as shown in Figures 5b and 5c).

\section{Concluding Remarks}

In this work, by considering an elastic medium with a screw dislocation within the cylindrical coordinates $\left( r,\varphi ,z\right) $, we have used a space-like screw dislocation as a topological defect and studied its effects on non-relativistic PDM Schr"{o}dinger particle. Where we have extended the method used in
\cite{14,44,45} to study PDM systems by using recent results of \cite{19,27} for the PDM kinetic operator and the PDM- momentum operator given by \eqref{eq:T} and \eqref{eq:P}, respectively.  We have divided this work into two parts. In the first part, we investigated the influence of the screw dislocation on the PDM particle for two configurations:  (i)  Quasi-free PDM Schr\"{o}dinger particles with space-like screw dislocation, and (ii) PDM Schr\"{o}dinger particles under the influence of space-like screw dislocation and interaction potential. In a simple one-dimensional textbook, purely radial Schr\"{o}dinger equations are obtained and transformed into a radial one-dimensional Schr\"{o}dinger form ( \eqref{eq:U4}, \eqref{eq:U6}) by using the PDM settings $f\left( r\right) =\lambda r^{\sigma}$. As a result of that, exact eigenvalues and eigenfunctions are obtained. We observed that the modification that happened in the energy spectrum is not only shifting the magnetic quantum number $|\ell -\chi k|$ (as happened in a constant-mass system ( \cite{14,44,45} ), but also the occurrence of the phenomenon of energy level crossings which is considered as a result of PDM setting ( as shown in Figures (1), (2), and (3a)). Such energy levels at crossing points of the quantum states involved in these plots (which are labeled as $\left( n_{r},\ell \right) $ for random quantum numbers $n_{r}$ and $\ell $) indicate that there could be more than one quantum state sharing the same energy at each crossing point ( i.e., occasional degeneracies). In addition, the effect of the screw dislocation with the PDM settings determined infinite degeneracies between some chosen quantum states (documented in Figures (1), (2), (3a)). However, the shifts in the energy levels that occurred due o the combined effects of the screw dislocation and the scalar potential produce a different pattern of the energy levels crossings (as shown in Fig.3a). Moreover, this potential reduced the occurrence of infinite degeneracies of energy levels. 

In the second part, we have considered PDM charged particles moving under not only the influence of the dislocation but also under the effect of magnetic and Aharonov-Bohm flux fields, for both cases: in the absence of potential energy and the presence of scalar potential energy. We have implemented the
PDM-minimal-coupling\ recipe \cite{48,49}, along with the PDM-momentum operator \cite{27}, and explored the separability of the problem under radial
cylindrical settings by using the same PDM settings, $f\left( r\right)=\lambda r^{\sigma }$, we reported on the exact solvability (both eigenvalues and eigenfunctions) of our PDM charged particles. This interaction produced (as in the previous part) a shift in the angular momentum which is $\left\vert \ell -\chi k-\frac{q\Phi }{2\pi }%
\right\vert $ because of the PDM-minimal-coupling between the momentum operator and the vector potential \eqref{eq:A1} which include the magnetic and Aharonov-Bohm flux fields. It also led to the occurrence of the phenomenon of energy level crossings, and occasional and infinite degeneracies of some quantum states mentioned in 
Fig.4, and Fig.5. Moreover, Fig.4 showed that the magnetic field $B_{\circ }$ and the harmonic part of the potential affected the energy spectrum by decreasing the values of the energies where most of them became negative. This is documented in section III.

Finally, although the topological defect of the screw dislocation modified the energy spectrum, the influence of the spatial dependence of the mass of
the system in the presence of external fields yields a new contribution to energy levels creating a set of new eigenvalues.

This study has investigated, for the first time, the influence of the topological defect of the screw dislocation on the non-relativistic PDM quantum particle
under the effect of external fields, and a new Hamiltonian was built suitable for studying such a system. Thus, this work opens new discussions regarding the position-dependent concept and different topological effects and provides a good starting point for future research.


\begin{thebibliography}{99}

\bibitem{1} T. W. B. Kibble, Phys. Rep. 67 (1980) 183.

\bibitem{2} A. Vilenkin, Phys. Rev D 23 (1981) 852.

\bibitem{3} A. Vilenkin, Phys. Rep. 121 (1985) 263.

\bibitem{4} E. A. F. Braggan\c{c}a, R. L. L.  Vit\'{o}ria, H. Belich, E. R. B. de Mello, Eur. Phys. J. C 80 (2020) 206.

\bibitem{5} R. A. Puntigam, H. H. Soleng, Class. Quant. Gravit. 14 (1997) 1129.

\bibitem{6} W. C. F. da Silva, K Bakke, R. L. L.  Vit\'{o}ria, Eur. Phys. J. C 79 (2019) 657.

\bibitem{7} R. L. L. Vit\'{o}ria, Eur. Phys. J. C 79 (2019) 844.

\bibitem{8} M. O. Katanaev, I. V. Volovich, Ann. Phys. 216 (1992) 1.

\bibitem{9} H. Kleinert, Gauge Fields in Condensed Matter Vols 1, 2, World Scientific, Singapore, 1989.

\bibitem{10} K.C. Valanis e, V. P. Panoskaltsis, Acta Mech. 175 (2005) 77.

\bibitem{11} B. Linet, Gen. Relat. Gravit. 17 (1985) 1109.

\bibitem{12} A. Vilenkin, Phys. Lett. B 133 (1983) 177.

\bibitem{13} A J Heeger, Comments Solid State Phys. 10 (1981) 53.

\bibitem{14} G. de A Marques, C Furtado, V B Bezerra, F Moraes, J. Phys. A: Math. Gen. 34 (2001) 5945.

\bibitem{15} M. O. Katanaev, 2005 Phys. Usp. 48 675.

\bibitem{16} L. Dantas , C. Furtado and A. L. Silva Netto, Phys. Lett. A 379 (2015) 11.

\bibitem{17} B. A. Bilby, R. Bullough and E. Smith, Proc. R. Soc. A 231(1955) 263.

\bibitem{18} P. M. Mathews, M. Lakshmanan, Quart. Appl. Math. 32 (1974) 215.

\bibitem{19} O. von Roos, Phys. Rev. B 27 (1983) 7547.

\bibitem{20} J. F. Cari\~{n}ena, M. F. Ra\~{n}ada, M. Santander, M. Senthilvelan, Nonlinearity 17 (2004) 1941.

\bibitem{21} O. Mustafa, J Phys A: Math. Theor.52 (2019) 148001.

\bibitem{22} O. Mustafa, Phys. Lett. A 384 (2020) 126265.

\bibitem{23} O. Mustafa, Euro. Phys. J. Plus 136 (2021) 249.

\bibitem{24} O. Mustafa, Phys. Scr. 95 (2020) 065214.

\bibitem{25} O. Mustafa, S. H. Mazharimousavi, Int. J. Theor. Phys 46 (2007) 1786.

\bibitem{26} Z. Algadhi, O. Mustafa, Ann. Phys. 418 (2020) 168185.

\bibitem{27} O. Mustafa, Z. Algadhi, Eur. Phys. J. Plus \textbf{134} (2019) 228.

\bibitem{28} A. Khlevniuk, V. Tymchyshyn, J. Math. Phys. 59 (2018) 082901.

\bibitem{29} O. Mustafa, J. Phys. A; Math. Theor. 48 (2015) 225206.

\bibitem{30} A. de Souza Dutra, C A S Almeida, Phys Lett. A 275 (2000) 25.

\bibitem{31} M. A. F. dos Santos, I. S. Gomez, B. G. da Costa, O. Mustafa, Eur. Phys. J. Plus 136 (2021) 96.

\bibitem{32} R. A. El-Nabulsi, Few-Body syst. 61 (2020) 37.

\bibitem{33} R. A. El-Nabulsi, J. Phys. Chem.Solids 140 (2020) 109384.

\bibitem{34} R. A. El-Nabulsi, Waranont Anukool, Applied Physic A 127 (2021) 856.

\bibitem{35} C. Quesne, J. Math. Phys. 56 (2015) 012903.

\bibitem{36} A. K. Tiwari, S. N. Pandey, M. Santhilvelan, M. Lakshmanan, J. Math. Phys. 54 (2013) 053506.

\bibitem{37} M.O. Katanaev, I.V. Volovich, Ann. Phys. (NY) 271 (1999) 203.

\bibitem{38} I.L. Shapiro, Phys. Rep. 357 (2002) 113.

\bibitem{39} F. Justo Joao, V.C. Assali Lucy, Physica B, 489 (2001) 308--310.

\bibitem{40} M.J. Bueno, C. Furtado, K. Bakke, Physica B 496 (2016) 45.

\bibitem{41} R.C.A. de Lima, C. Furtado, F. Moraes, Europhys. Lett. 62 (2003) 306.

\bibitem{42} C. Furtado, F. Moraes, J. Phys. A: Math. Gen. 33 5(2000) 513.

\bibitem{43} K. Bakke, Phys B: Condensed Matter 537 (2018) 346-348.

\bibitem{44} M.J. Bueno, C. Furtado, K. Bakke, Physica B 496 (2016) 45.

\bibitem{45} C. Furtado, F. Moraes, Europhys. Lett, 45 (1999) 279-282.

\bibitem{46} N. Soheibi, M. Hamzavi, M. Eshghi, S. M. Ikhdair, Eur. Phys. J. B 90 (2017) 212.

\bibitem{47} S. Azevedo, Phys. Lett A 306 (2002) 21-24.

\bibitem{48} O. Mustafa, Z. Algadhi, Eur. Phys. J. Plus 135 (2020) 559.

\bibitem{49} O. Mustafa, Z. Algadhi, Chin. J. Phys. 65 (2020) 554.

\bibitem{50} O. Mustafa, Ann. Phys. 440 (2022) 168857.

\bibitem{51} O. Mustafa, Eur. Phys. J. C. 82 (2022) 82.

\bibitem{52} O Mustafa , Confined Klein-Gordon oscillators in Minkowski spacetime and a pseudo-Minkowski spacetime with a space-like dislocation: PDM KG-oscillators, isospectrality and invariance", arXiv:2111.10077, Ann. Phys. (2022) in press.

\bibitem{53} S. L.A. Netto, C. Furtado, J. Phys.: Condens. Matter, 20 (2008) 125209.

\bibitem{54} S. Gasirowicz,  Quantum Mechanics, 3rd Edition ( John Wiley and Sons, Inc., New Jersey 2003).

\bibitem{55} E.R.Arriola, A.Zarzo, J.S.Dehesa, J. Comput. Appl. Math. 37 (1991) 161-169.

\bibitem{56} A.Ronveaux, Heun's Differential Equation, (Oxford University Press, Oxford, 1995).

\end{thebibliography}
\end{document}